\documentclass[11pt,twoside]{article}
\usepackage{newpasp}
\usepackage{epsf}
\begin{document}

\title{Observations and models of the embedded phase of high-mass star
  formation}
\author{Floris F.\ S.\ van der Tak}
\affil{Max-Planck-Institut f\"ur Radioastronomie, Auf dem H\"ugel 69,
  53121 Bonn, Germany}

\begin{abstract}
  This paper is a review and an update on recent work on the physical
  and chemical structure of the envelopes of newly born massive stars,
  at the stages preceding ultracompact H~{\sc II} regions.
  It discusses methods and results to determine total mass,
  temperature and density structure, ionization rate, and
  depth-dependent chemical composition.
\end{abstract}

\section{Introduction}
The first major question in the study of high-mass star formation is
whether or not the process is a scaled-up version of the formation of
low-mass stars, which are known to form by accretion via a disk, and
to disperse their natal envelopes through bipolar outflows. At the other
extreme, high-mass stars may form by the coagulation of lower-mass
stars or protostellar cores 
%(see reviews by Churchwell 1999 and Kurtz et al.\ 2000). 
(see reviews by Churchwell and by Evans in this volume).
The second major question of high-mass star formation
is its relation to clustered star formation: does the stellar density
influence the emergent mass spectrum? To what degree do the conditions
in the initial molecular cloud determine the properties of the stellar
population it produces? 
%\nocite{kurt00,chur99}

The study of high-mass star formation is hampered by the short
time-scale of the process. For example, a late O-type star lasts only
$\sim$1~Myr, equal to the pre-main sequence lifetime of a 1~M$_\odot$
star. Moreover, 15\% of this time is spent embedded in natal gas and
dust that hides the photosphere from optical view. As a consequence,
known regions of high-mass star formation lie at distances of several
kpc, so that details $\la 1000$~AU are unresolved at typical $1''$
resolution. The large distances cause considerable confusion in these
regions, and make it difficult to construct source samples with a
common luminosity and distance, necessary for systematic studies.

To answer these questions, it is necessary to build up a statistically
significant sample of regions forming high-mass stars, and to study
their physical and chemical properties. This review describes recent
studies of molecular material in which stellar groups including
high-mass stars are forming. The focus is on detailed studies of small
samples; larger source samples are described by Bronfman et al.\ 
(1996), Brand et al.\ (2001) and Sridharan et al.\ (2001). The first
sample consists of nine deeply embedded objects of $L=10^{4}-10^{5}$
L$_\odot$, $d=1-4$~kpc, which have strong mid-infrared and weak
cm-wave emission.  The second set contains nine molecular clumps which
have $L=10^{5}-10^{6}$ L$_\odot$, $d=4-14$~kpc and weak mid-infrared
emission, and are close to ultracompact H~II regions. Both samples are
characterized by strong submillimeter continuum and line emission,
associated H$_2$O and CH$_3$OH masers, and molecular outflows. In many
cases, the distances are only kinematic estimates. Better distance
estimates may come soon from near-infrared spectroscopy (Hanson et
al.\ 1997; Kaper, this volume).
%\nocite{bron96} \nocite{brand01} \nocite{srid01}  \nocite{hans97}

\section{Physical structure}
The determination of the physical structure of high-mass star-forming
regions consists of two parts: first the measurement of the total mass
and the temperature structure, and second the determination of the
density structure. 

\begin{figure}
%  \plotone{gl2591_lnu.eps}
\plotfiddle{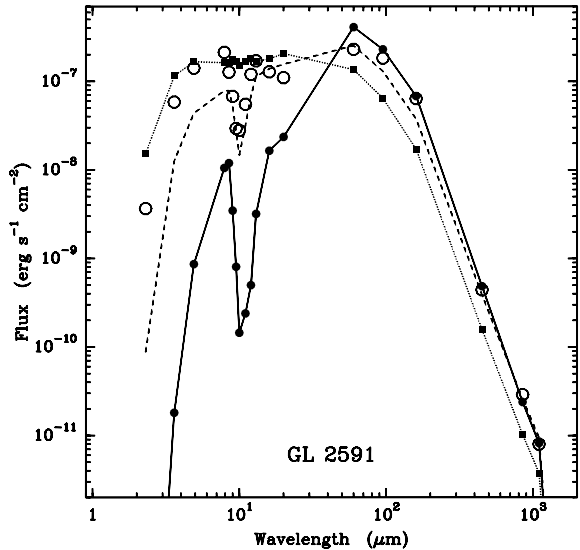}{6.5cm}{0}{43}{43}{-120}{-20} 
  \caption{Spectral energy distribution of GL~2591. Open circles: data;
    solid line: model with coagulated dust grains and ice mantles;
    dashed line: model with coagulated dust grains but no ice mantles;
    dotted line: model with lower optical depth to fit the
    near-infrared. Based on: van der Tak et al.\ (1999). }
  \label{fig:gl2591}
\end{figure}

\subsection{Mass and temperature structure}
\label{sec:temp}
The mass and temperature structure can be obtained by fitting dust
models to the broad-band infrared spectrum. Masses derived from
molecular line data are less reliable due to uncertain molecular
abundances.  As illustrated in Fig.~\ref{fig:gl2591} (solid line), a
single model cannot fit the entire spectrum, so that to measure the
mass, the model must be fitted to the optically thin part, at $\lambda
\ga 100\,\mu$m. Early models (Wolfire \& Cassinelli 1986; Churchwell
et al.\ 1990) mostly fitted the near-infrared
part of the spectrum, which indicates $\sim$5~times lower optical
depths than the submillimeter part due to deviations from spherical
geometry. This result holds for both bright and weak mid-infrared
sources (Harvey et al.~2000; Hatchell et al.~2000). 
%\nocite{wolf86,cww90} \nocite{harv00} \nocite{hatch00}

The source GL~2591 offers a rare opportunity to measure the
submillimeter dust opacity by comparing dust masses to those derived
from CO isotopic emission. While the CO mass is usually a lower limit
since an unknown fraction of CO is frozen out onto dust grains,
mid-infrared observations of GL~2591 indicate a ratio of gas-phase to
solid-state CO column density of $>400$ (Mitchell et al.\ 1989),
reducing the uncertainty in CO abundance to better
than a factor of two. 
%\nocite{mitc89} 

The dust masses of GL~2591 quoted in van der Tak et al.\ (1999) for
Draine and Lee (1984) and Mathis et al.\ (1983) dust are a factor of~8
too high due to the use of a grain size of $0.05\,\mu$m rather than
$0.1\,\mu$m.  However, the conclusion that the dust model of Ossenkopf
\& Henning (1994) matches the mass derived from C$^{17}$O best is
still valid. This model for dust in star-forming regions includes
grain coagulation and the formation of ice mantles.
%\nocite{fvdt99} \nocite{drain84} \nocite{math83} \nocite{ossen94} 

The derived temperature profile indicates $T$ $>$100~K at $r$ $<$1000~AU
from the central star. Within this radius, ice mantles on the grains
should have evaporated. The dotted line in Fig.~\ref{fig:gl2591} is
a calculation using the coagulated grain model without ice mantles.
This model clearly fits the mid-infrared part of the spectrum better
than the model with ice mantles.

\subsection{Density structure}
\label{sec:dens}
The density structure can be derived either by fitting dust models to
the spatial profile of dust continuum emission, or by fitting the line
spectrum of a molecule with a high dipole moment (such as CS) that is
sensitive to density. The first approach assumes that the grain
optical properties do not vary with position; the second that the
molecular abundance does not. While either assumption may be invalid,
the methods can serve as tests of each other.

Van der Tak et al.\ (2000a) found the two methods to
agree well for six embedded high-mass star-forming regions, while for
a seventh, S~140, the assumption of a centrally heated envelope
appeared invalid. The good agreement between the dust and CS results
may have been fortuitous due to the use of a Gaussian beam shape.
Models of the 350~$\mu$m dust emission profile of GL~2591 using the
actually measured beam shape indicate an $r^{-2}$ density profile,
steeper than the $r^{-1}$ profile found from CS (Mueller et al., this
volume). If this result also holds for the SCUBA data and the other
sources, it indicates a depletion of CS at small radii. The
alternative explanation of enhanced dust emissivity at small radii is
not predicted by dust models, in which the main effect would be ice
mantle evaporation, which drops the opacity.
%\nocite{fvdt00} 

\begin{figure}
\plotfiddle{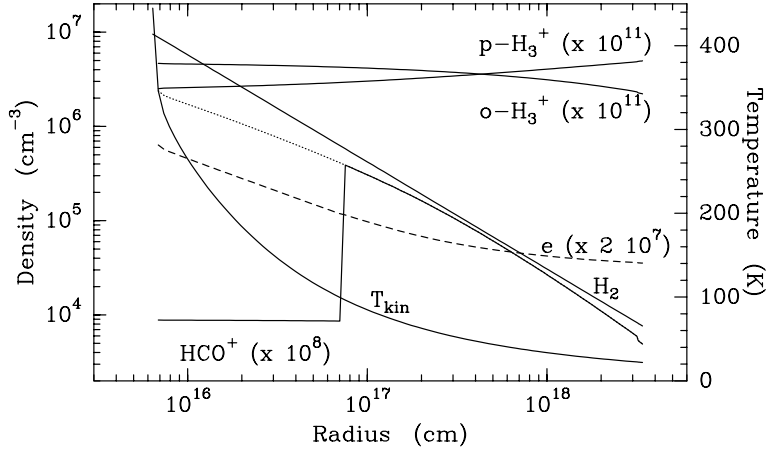}{5cm}{-90}{43}{43}{-150}{220} 
%  \plotone{zeta.eps}
  \caption{Density, temperature and ionization structure of
    GL~2136. The HCO$^+$ abundance drops at $T$=100~K due to reaction
    with evaporating H$_2$O ice. Based on van der Tak \& van Dishoeck (2000).}
  \label{fig:zeta}
\end{figure}

\section{Ionization}
\label{sec:ion}
The fractional ionization of molecular clouds controls both the
influence of magnetic fields on their dynamics and their gas-phase
chemistry. In regions shielded from ultraviolet starlight, the main
source of ionization is by cosmic rays which produce H$^+_3$, followed
by proton transfer reactions into HCO$^+$ and H$_3$O$^+$. Current best
estimates of the cosmic-ray ionization rate $\zeta$ based on the
abundances of OH and HD measured in visual absorption toward nearby
diffuse clouds have a factor of ten uncertainty (Federman et al.\ 
1996).  An alternative approach using the abundances of HCO$^+$ and
DCO$^+$ measured through millimeter emission lines towards nearby dark
clouds is no more accurate due to uncertain and potentially large
depletion of CO and O (Caselli et al.\ 1998; Williams et al.\ 1998).
At the current sensitivity, both these methods are limited to the
solar neighbourhood and cannot give information on variations of
$\zeta$ on Galactic scales.
%\nocite{feder96} \nocite{john98} \cite{casel98b} 

The infrared H$^+_3$ detections (Geballe \& Oka 1996; McCall et al.\ 
1999) for well-characterized lines of sight make the embedded
high-mass objects good targets for a more direct determination of
$\zeta$.  Van der Tak \& van Dishoeck (2000) combine the temperature
and density structure of seven embedded sources with a small chemical
network to compute their ionization structure (Fig.~\ref{fig:zeta})
and fitted the observed H$^+_3$ absorption and H$^{13}$CO$^+$ emission
lines.
%$\log(\zeta)=(-16.71 \pm ...)$ (s-1). 
The mean value of $\zeta = 2.6 \times 10^{-17}$~s$^{-1}$ is consistent
with the diffuse and dark cloud estimates, and with recent Voyager and
Pioneer spacecraft data at distances of up to 60~AU from the Sun
(Webber 1998). The source-to-source spread in $\zeta$ of a factor of
2-3 on scales of a few kpc agrees with results from $\gamma-$ray
observations (Hunter et al.\ 1997). This agreement may be a
coincidence since the $\gamma-$rays are of higher energy ($\sim$GeV)
than the particles providing the bulk of the ionization (a few
10~MeV). Intervening clouds affect the H$^+_3$ but not the
H$^{13}$CO$^+$ data.
%\nocite{bmcc99} \nocite{gebal96} \nocite{zeta00} \nocite{webb98}  \nocite{hunt97}

Alternatively, the spread may be due to shielding against cosmic rays.
The data suggest a trend of decreasing $\zeta$ with increasing total
column density, but not at a significant level.  However, HCO$^+$
probes the outer envelopes (Fig.~\ref{fig:zeta}), because in regions
with $T>100$~K, HCO$^+$ reacts with evaporating H$_2$O ice.  Using
tracers of the innermost parts of a low-mass core, Caselli et al.\ 
(2001) found $\zeta =6 \times 10^{-18}$~s$^{-1}$.  Follow-up studies
may reveal by how much cosmic rays are attenuated inside dense cores.
%\nocite{casel01b} 

\section{Chemical stratification}
\begin{figure}
\plotfiddle{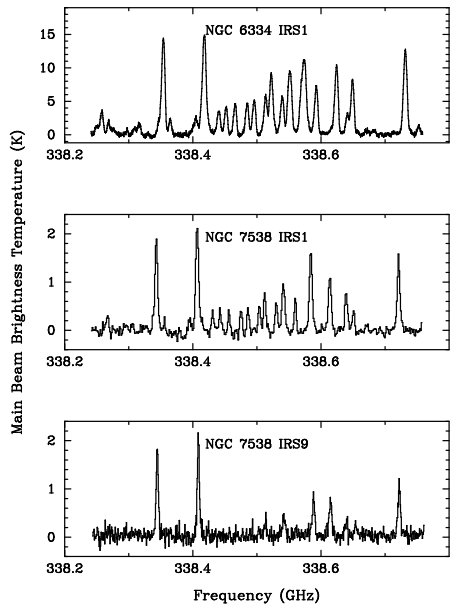}{7cm}{0}{43}{43}{-120}{-60} 
%  \plotone{ch3ohdata.ps}  
  \caption{Spectra of the J=7-6 transition of CH$_3$OH at 338 GHz,
    taken with the JCMT, toward a `cold', a `jump' and a `warm' source
    (bottom to top).}
  \label{fig:ch3oh}
\end{figure}

The models described in \S~2 have allowed us to study the chemical
composition of the envelopes of embedded high-mass stars as a function
of distance from the central star.  The first example is the CH$_3$OH
molecule studied by van der Tak et al.\ (2000b).  Modeling of spectra
indicates three types of sources (Fig.~\ref{fig:ch3oh}): `cold'
sources with CH$_3$OH/H$_2$ $\sim 10^{-9}$, `warm' sources with
CH$_3$OH/H$_2$ $\sim 10^{-8}$, and sources where the abundance `jumps'
from $\sim 10^{-9}$ to $\sim 10^{-7}$. This `jump' occurs at $T\approx
100$~K, where solid H$_2$O, the bulk of icy grain mantles, evaporates.
Combined with the high observed abundances of solid CH$_3$OH in these
sources, these `jumps' suggest an evolutionary sequence where CH$_3$OH
forms on dust grains in `cold' sources, then evaporates into the gas
phase in `jump' sources, and is broken down by reactions with ions in
`warm' sources.  However, gas-phase CH$_3$OH abundances measured in
the submillimeter are factors of $\ga$100 lower than those derived
from solid-state data. Since the nearby Orion region is an exception,
resolution could play a role.  Observations of gas-phase CH$_3$OH in
infrared absorption with ground-based instruments with high spectral
resolution may clarify this point.
%\nocite{meth00}

A much earlier phase of evolution is probed by the CO molecule.
Abundances derived from C$^{17}$O line emission increase
systematically by a factor of $\approx$3 as
$\bar{T}$=20$\rightarrow$40~K, where $\bar{T}$ is the average
temperature of the envelope, weighted by mass.  Since these
temperatures are the sublimation points of pure CO and mixed
CO--H$_2$O ices, freeze-out and evaporation appear to control the CO
abundance (van der Tak et al.\ 2000a). In low-mass cores, Kramer et
al.\ (1999) found a similar relation of CO abundance with temperature.
A third example of depth-dependent chemistry is the enhancement of HCN
at temperatures $\ga 300$~K (Boonman et al.\ 2001).  On the other
hand, the abundance of H$_2$CO was found to be constant within the
range $T=20-240$~K, suggesting that this species is formed either in
the gas phase or in CO-dominated ice. The latter idea is consistent
with evidence from D$_2$CO (Ceccarelli et al.\ 2001).
%\nocite{kram99} \nocite{cecc01} \nocite{boon01}

The CO$_2$ molecule may be tracing the thermal history of star-forming
regions. While the CO$_2$ ice abundance is high and evaporation of
ices is observed, its gas-phase abundance is low, so that CO$_2$ must
be rapidly destroyed after it evaporates (van Dishoeck 1998).
However, there is no obvious low-temperature destruction path.  The
required $T\ga 800$~K for CO$_2$ destruction is well above measured
values, suggesting impulsive heating events in the past.
Possibilities include outflow shocks (Charnley \& Kaufman 2000), if
desorption of the ices is also caused by these shocks.  Other options
are episodic accretion (Doty et al.\ 2001a) and the infrared flares
discussed by Bally (this volume).
%\nocite{evd98b} \nocite{char00}\nocite{doty01a}

\begin{figure}
  \begin{center}
\plotfiddle{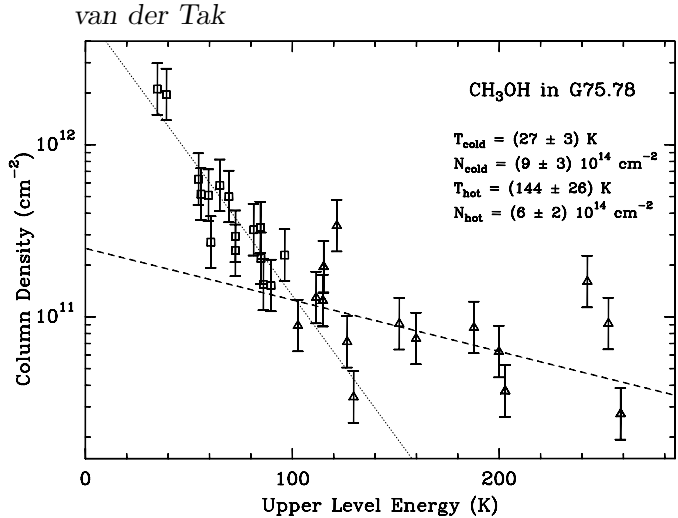}{5.5cm}{-90}{35}{35}{-150}{200}     
    \caption{Rotation diagram of CH$_3$OH toward G75.78, a hot core
      close to an ultracompact H~{\sc II} region.}
    \label{fig:g7578}
  \end{center}
\end{figure}

\section{Envelope evolution}
The evolution of the temperature and density structure of embedded
objects was studied by van der Tak et al.\ (2000a). 
Conventional tracers of physical evolution, used for low-mass objects,
do not change much over the narrow age range of the objects studied in
detail. In particular, the shape of the spectral energy distribution
does not change with time, as most material remains at low
temperatures. The radio continuum flux does not seem indicative
either, partly due to the unknown contribution of shocks, and partly
because the surface temperatures of the ionizing stars are not well
known.
%\nocite{fvdt00}

However, the embedded objects can be ordered in a temperature
sequence, based on the gas/solid ratios of H$_2$O and CO$_2$, the
fraction of heated solid $^{13}$CO$_2$, the %far-infrared
45/100~$\mu$m colour, and
the excitation temperatures of CO, C$_2$H$_2$ and CH$_3$OH. These
indicators cover a wide range of temperatures and spatial scales,
demonstrating that geometry is not an effect.  Since crystallization
is an irreversible process, the heated $^{13}$CO$_2$ ice measures the
maximum temperature. Hence, the temperature change must be a
systematic increase rather than random fluctuations, although
gas-phase CO$_2$ argues the opposite.  Still, the sequence may be
evolutionary since it is correlated with the ratio of envelope mass to
bolometric luminosity. If confirmed through a larger source sample,
this correlation indicates the progressive dispersal of the envelope.
%by the star(s) it produced.

%However, this scenario needs to be tested through other species. For
%example, episodic heating is not suggested by the correlation of the
%fraction of heated $^{13}$CO$_2$ ice with other temperature indicators
%(Boogert et al.\ 2000).  \nocite{boog00}

As for the time order of particular phenomena, the small body of
existing data suggests that CH$_3$OH masers occur first, followed by
H$_2$O masers, and later by OH masers. The situation is complex as
each kind of maser may trace more than one physical component (Minier
et al.\ 2001). Very massive and energetic outflows occur throughout
the embedded phase, like for low-mass objects (Beuther, this volume).
%\nocite{minier01} 

The chemical composition of massive protostellar envelopes may serve
as an independent probe of their structure and evolution. A good test
case is GL~2591, for which a large body of data has been analyzed,
both in submillimeter emission (JCMT, SWAS) and in infrared absorption
(ISO-SWS, IRTF).  Doty et al.\ (2001b) coupled the temperature and
density structure of GL~2591 by van der Tak et al.\ (2000) to a
complete gas-phase chemical network. It appears that the measured
molecular abundances can be reproduced to within factors of a few for
a source age of $\sim 3 \times 10^4$~yr.
%\nocite{doty01b} 

\section{Molecular gas near ultracompact H II regions}
\label{sec:uchii}
The results discussed so far pertain to a limited range of ages and
luminosities. To see just how limited, we are investigating the set of
line-rich molecular line sources near ultracompact H~{\sc II} regions
from Hatchell et al.\ (1998).  Observations of CH$_3$OH
(Fig.~\ref{fig:g7578}) indicate the presence of both cold (20-50~K)
and warm (110-500~K) gas. The column density ratio of these components
ranges from $<$0.1 to $>$60, comparable to the values found in the
embedded bright mid-infrared sources. The large implied masses of cold
gas argue against a more evolved status of these regions, as would
seem plausible based on their higher abundances of complex organic
molecules.  To analyze the data in more detail, we are developing
models for the temperature and density structure, based on tracers of
dust (SCUBA) and gas (CS lines).

\section{Conclusions}
Regions forming high-mass and low-mass stars have a different
observational appearance, but their underlying structure and evolution
seem similar.  Regions forming massive stars are larger and hence more
massive, but their absolute densities and their density distributions
are similar to those found in low-mass star-forming regions. The more
rapid dispersal of high-mass envelopes may be mainly driven by the
faster pace of evolution of the central stars. There is mounting
evidence for disks around high-mass stars, outflows are found to be
ubiquitous, and compact dust and gas cores seem common as well.
%Thus, at this point, the formation processes of high-mass and
%low-mass stars do not appear fundamentally different.

The source samples with and without nearby ultracompact H~II regions
appear to evolve in parallel, suggesting that the proximity of radio
emission and the shape of the spectral energy distribution are not
good clocks. However, the excitation and abundance of selected
molecules appear excellent tracers of evolution.

\acknowledgments
The author thanks Neal Evans and Ewine van Dishoeck for comments on
this paper.

%\reference
%{Boogert}, A. C.~A., {Ehrenfreund}, P., {Gerakines}, P.~A., et al.\
%%{Tielens}, A. G. G.~M., {Whittet}, D. C.~B., {Schutte}, W.~A., {van
%%  Dishoeck}, E.~F., {de Graauw}, T., {Decin}, L., \& {Prusti}, T. 
%2000, \aap, 353, 349
%

\end{document}